\documentclass[journal]{IEEEtran}
\makeatletter
\newif\if@restonecol
\makeatother

\usepackage{lineno,hyperref}
\usepackage[english]{babel}
\usepackage{amsmath}
\usepackage{bm}
\usepackage{graphicx}
\usepackage{tabularx}
\usepackage{multirow}
\usepackage{multicol}
\usepackage{caption}
\usepackage{float}
\usepackage{cite}
\usepackage{epstopdf}
\usepackage[ruled,linesnumbered]{algorithm2e}
\usepackage{color}
\usepackage{colortbl}

\modulolinenumbers[5]

\ifCLASSINFOpdf
\else
\fi

\begin{document}
\captionsetup[figure]{name={Fig.}}
%
\title{GA Based Q-Attack on Community Detection}

\author{Jinyin Chen, Lihong Chen, Yixian Chen, Minghao Zhao, Shanqing Yu, Qi Xuan,~\IEEEmembership{Member, IEEE}, Xiaoniu Yang
\thanks{This work was partially supported by National Natural Science Foundation of China (61502423, 61572439), Zhejiang Provincial Natural Science Foundation of China (LR19F030001, LY19F020025), Zhejiang Science and Technology Plan Project (LGF18F030009), and the Key Technologies, System and Application of Cyberspace Big Search, Major Project of Zhejiang Lab (2019DH0ZX01). \emph{(Corresponding author: Qi Xuan.)}}
\thanks{J. Chen, S. Yu, and Q. Xuan are with the Institute of Cyberspace Security, and the College of Information Engineering, Zhejiang University of Technology, Hangzhou 310023, China; and also with the Zhejiang Lab, Hangzhou 311121, China (e-mail: chenjinyin@zjut.edu.cn; yushanqing@zjut.edu.cn; xuanqi@zjut.edu.cn).}
\thanks{L. Chen, Y. Chen, and M. Zhao are with the Institute of Cyberspace Security, and the College of Information Engineering, Zhejiang University of Technology, Hangzhou 310023, China (e-mail: coffeeclh@163.com; yichixchen@163.com; yzbyzmh1314@163.com).}
\thanks{X. Yang is with the Institute of Cyberspace Security, Zhejiang University of Technology, Hangzhou 310023, China, and with the Science and Technology on Communication Information Security Control Laboratory, Jiaxing 314033, China (e-mail: yxn2117@126.com).}}


\maketitle

\begin{abstract}
Community detection plays an important role in social networks, since it can help to naturally divide the network into smaller parts so as to simplify network analysis. However, on the other hand, it arises the concern that individual information may be over-mined, and the concept community deception has been proposed to protect individual privacy on social networks. Here, we introduce and formalize the problem of \emph{community detection attack} and develop efficient strategies to attack community detection algorithms by rewiring a small number of connections, leading to privacy protection. In particular, we first give two heuristic attack strategies, i.e., Community Detection Attack (CDA) and Degree Based Attack (DBA), as baselines, utilizing the information of detected community structure and node degree, respectively. And then we propose an attack strategy called ``Genetic Algorithm (GA) based Q-Attack", where the modularity $Q$ is used to design the fitness function. We launch community detection attack based on the above three strategies against six community detection algorithms on several social networks. By comparison, our Q-Attack method achieves much better attack effects than CDA and DBA, in terms of the larger reduction of both modularity $Q$ and Normalized Mutual Information (NMI). Besides, we further take transferability tests and find that adversarial networks obtained by Q-Attack on a specific community detection algorithm also show considerable attack effects while generalized to other algorithms.
\end{abstract}


\IEEEpeerreviewmaketitle

\section{Introduction}\label{Section1}
%
%
%
%
\IEEEPARstart{C}{omplex} networks can well represent various complex systems in our daily life, such as social networks~\cite{xuan2018social,xuan2015temporal}, biological networks~\cite{garcia2018applications}, power networks~\cite{chen2018robustness} and financial networks~\cite{schiavo2010international}. Network science has been a very active field because of its highly interdisciplinary nature~\cite{lancichinetti2009community}, problems such as link prediction~\cite{fu2018link}, node classification~\cite{kipf2016semi}, network reconstruction~\cite{zhang2018reconstructing} or community detection~\cite{newman2003structure} have been widely studied. And as one hot topic in this field, community detection has attracted great attentions of researchers from various disciplines~\cite{newman2001structure,gleiser2003community,abbasi2007survey}. Detecting communities in the network has played an important role in getting a deep understanding of its organizations and functions.

However, with the rapid development of community detection algorithms in recent years~\cite{fortunato2016community}, a new challenging problem arises: information over-mined. People realize that some information, even their own privacy, is going to be over-mined by those social network analysis tools. The web we browse, the people we contact with and more things like that are all recorded through the Internet. Our interests, circle of friends, partners on business can be easily detected once these data are utilized~\cite{xuan2018modern,fu2018link}. Community hiding or deception~\cite{nagaraja2010impact,waniek2018hiding,fionda2018community} is put forward to hide certain community. Still, community detection attack is launched on the network to change some of its structure thereby making the performance of related algorithms worse and the accuracy of results lower.

The problem of making communities in the network more difficult to detect through attack strategies against detection algorithm is worthy studying because the good attack strategies can be a guideline for individuals or organizations to disguise their social circles and change the way they interact with others, lest some privacies get leaked~\cite{waniek2018hiding}. In this paper we focus on investigating the attack strategy and testing the effectiveness of these strategies through experiments. In particular, we propose two heuristic attack strategies to address the problem shown in the last paragraph. Heuristic strategy is vital because it is easy to implement and plays well sometimes. During the design process of heuristic attack strategies, we take community detection algorithms and some properties of complex networks into consideration. One question is about what kind of nodes to be chosen as targets may benefit the attack effect. Typically, node centrality is a common-used index to measure the importance of nodes in the network. It mainly includes degree, betweenness and closeness. Compared with betweenness, whose calculation desires global information, degree is a quantity only based on local information with less computation~\cite{holme2002attack}. Thus, we propose an attack strategy based on degree and take some tests.

Furthermore, we consider the design of attack strategy as an optimization problem. Owing to the good performance in solving complex optimization problems, evolutionary algorithms, such as Genetic Algorithm (GA), have been widely used in various disciplines, including complex networks~\cite{tasgin2007community,liu2011application,wang2012community}. Searching for attack schemes which make the performance of community detection algorithms decrease most under given attack cost is a typical combinatorial optimization problem. We propose an attack strategy based on GA, namely \emph{Q-Attack}, where the modularity $Q$ is used to design the fitness function, and show its effectiveness on different community detection algorithms and several real-world networks. The major contributions of our work are summarized as:
\begin{itemize}
\item First, we introduce and formalize community detection attack problem, which is to implement global community deception and privacy protection;
\item Second, we propose two heuristic attack strategies and GA based Q-Attack method to launch attack with negligible rewiring to cheat community detection algorithms;
\item Third, we conduct comprehensive experiments to compare the attack effects of different attack strategies against six community detection algorithms on several real-world networks;
\item Finally, the transferability  of Q-Attack is validated and we think that community detection attack can also be a robust evaluation metric for anti-attack capacity comparison of community detection algorithms.
\end{itemize}

The rest of paper is organized as follows. Sec.~\ref{section2} reviews the related work on community detection and attacks on complex network. Sec.~\ref{section3} presents our attack strategies for destroying community structure in the network, including heuristics and optimization methods. We demonstrate the experiments and results in Sec.~\ref{section4} and finally draw conclusions in Sec.~\ref{section5}.

\section{Related work}
\label{section2}
\subsection{Community Detection Algorithms}
\label{section2.1}
A large number of community detection algorithms have been proposed since the problem brought up~\cite{fortunato2016community}. And detection algorithms continue to spring up because of the diversity of networks in our real world. Here we mainly give a brief introduction of some ones.

Girvan and Newman~\cite{girvan2002community} first proposed a community detection algorithm based on the betweenness of edges and experiments showed that it did a good job in small networks. Some researchers made changes on this study and proposed new ones, e.g.,Tyler et al.~\cite{tyler2005mail} got approximate betweenness using the fast algorithm of Brandes, Radicchi et al.~\cite{radicchi2004defining} used clustering coefficient of edges instead so that only local information was required and the efficiency of the algorithm was improved. Besides, Newman~\cite{newman2004fast} also proposed a greedy algorithm which mainly aimed at optimizing modularity. CNM was an improvement of Newman's method, which was proposed by Clauset et al.~\cite{clauset2004finding}. The main technology used in this algorithm was the data structures that called stacks. Futhermore, Blondel et al.~\cite{blondel2008fast} divided the modularity optimization problem into two levels and the computational complexity was essential linearly. Spectral clustering algorithms~\cite{newman2006modularity,newman2006finding} leveraged the eigenvalue spectrum of several graph matrices to detect the community structure in networks.

Except these methods based on modularity optimization, researchers also considered the problem of community detection from other perspectives. Label Propagation Algorithm (LPA) was proposed by Raghavan et al.~\cite{raghavan2007near}, whose main idea was updating the label of each node according to its neighbors. Rosvall et al.~\cite{rosvall2008maps} proposed an algorithm called Infomap, which was based on information theory. This algorithm found out the partition for the network through optimizing the average length of description $L(M)$. Further, Ronhovde and Nussinov~\cite{ronhovde2009multiresolution} proposed a community detection algorithm based on the minimization of the Hamiltonian of Potts model.

\subsection{Attacks on Complex Networks}
\label{section2.2}
Several earlier researches have studied the influence of attacks on the network performance. Holme et al.~\cite{holme2002attack} proposed four strategies to investigate the attack vulnerability of different kinds of networks, including \emph{ID removal}, \emph{IB removal}, \emph{RD removal} and \emph{RB removal}, both for nodes and links. ID removal represented removing nodes (or links) according to the descending order of degrees in the initial network while ``B" represented betweenness and ``R" represented recalculation. In this study, it was validated that the network structure changed and some performances degraded via attacks. Bellingeri et al.~\cite{bellingeri2014efficiency} found that node deletion according to betweenness centrality was most efficient while using the size of Largest Connected Component (LCC) as a measure of network damage. As for the community structure, Karrer et al.~\cite{karrer2008robustness} studied the robustness of community structure in networks through perturbing networks in a specific way. They constructed a random network with the same number of nodes and links as the original one, then replaced the links in the original network with that in the random one with the probability of $\alpha$. These studies shed lights on the further work about taking attacks on networks.

Adversarial attacks against some network algorithms are emerging in recent years. Yu et al.~\cite{yu2018target} developed both heuristic and evolutionary methods to add subtle perturbations to original networks, so that some sensitive links would be harder to predict. Dai et al.~\cite{dai2018adversarial} proposed three methods to modify graph structure, including \emph{RL-S2V}, \emph{GradArgmax} and \emph{GeneticAlg}, aiming at fooling the classifiers based on GNN models. Moreover, Z\"{u}gner et al.~\cite{zugner2018adversarial} proposed the first adversarial attack on networks using GCN, namely \emph{NETTACK}, and found that this method was effective on node classification task. Quite recently, Bojchevski and G\"{u}nnemann~\cite{bojchevski2018adversarial}, as well as Chen et al.~\cite{chen2018fast}, proposed adversarial attacks on network embedding at the same time, which may promote the extensive discussion on the robustness of network algorithms, since network embedding establishes a bridge between network space and Euclidean space and thus facilitates many downstream algorithms, such as node classification and link prediction.

\begin{figure*}[!t]
\centering
\includegraphics[width=\linewidth]{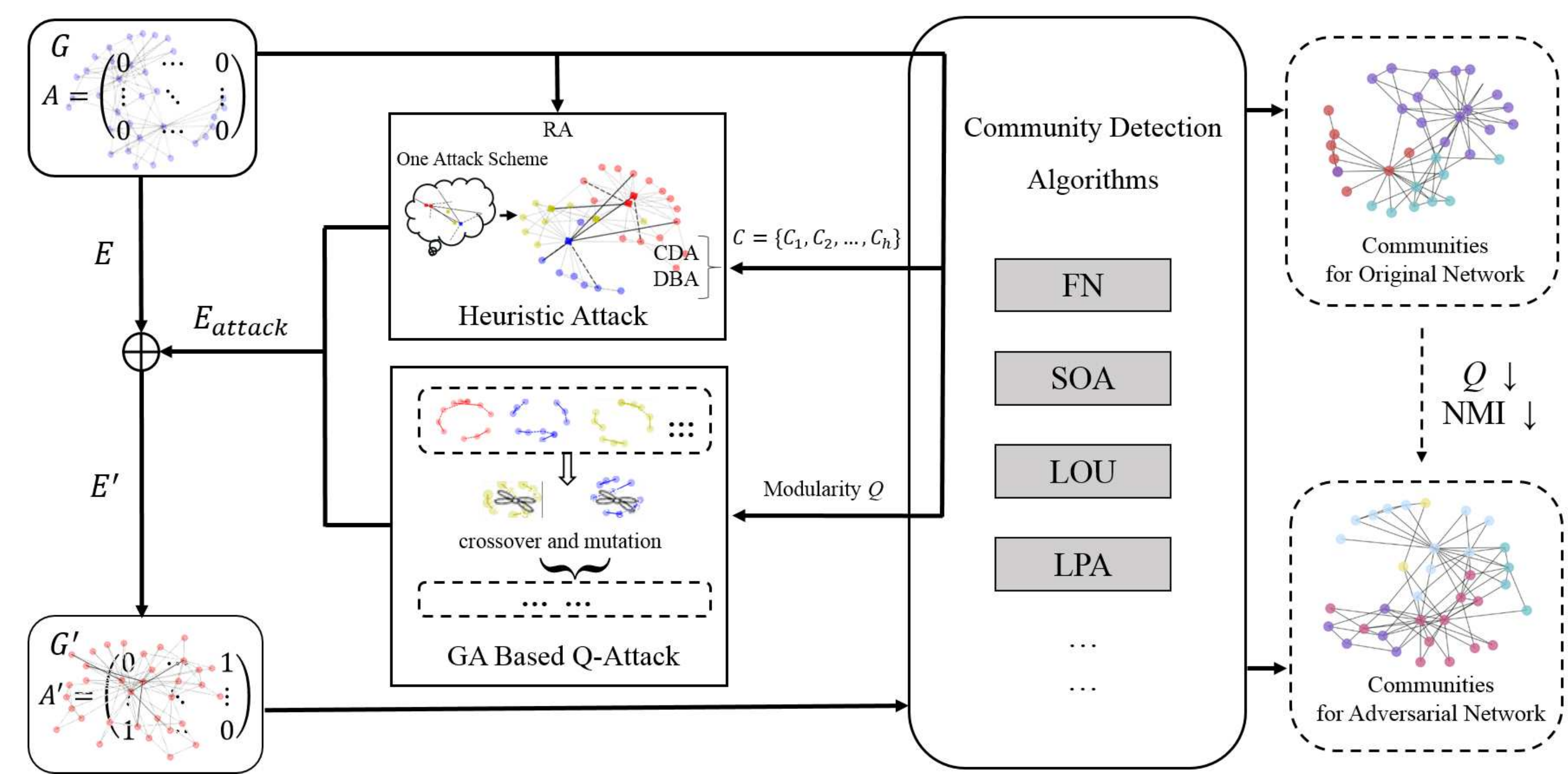}
\caption{The framework of community detection attack. We get the attack scheme $E_{attack}$ combined with community detection algorithms. Two metrics $Q$ and NMI are used to evaluate the attack effect.}
\label{fig:1}
\end{figure*}

\subsection{Researches on Anti-Detection of Community}
\label{section2.3}
As for privacy concerns aroused by community detection, some researchers started paying attention to this problem during the last two years. Waniek et al.~\cite{waniek2018hiding} came up with two heuristic algorithms ROAM and DICE for hiding individuals and communities, respectively. DICE implemented through disconnecting $d$ intra-links of given community $C$ and connecting $b-d$ inter-links under budget $b$ and a measure was proposed to evaluate the concealment. Finoda et al.~\cite{fionda2018community} proposed the concept of community deception to cope with community detection. Safeness-based deception algorithm and modularity-based deception algorithm were mainly tested through amounts of experiments. Another job they did was defining the deception score to evaluate effects of proposed methods. Similar to the measure proposed in~\cite{waniek2018hiding}, deception score also took community spread and hiding into account and it was more comprehensive while reachability was also considered. Extensive distribution of members in $C$ and less percentages of $C^{'}s$ for each $C_i\in \overline{C} $ showed a good hiding, where $\overline{C}$ was the community structure detected by specific community detection algorithm and $C_i$ was one element in $\overline{C}$.

There is still little attention that has been paid on the problem of taking attacks on the network to make it harder for algorithms to detect the community structure. And both Waniek's and Finoda's researches are dedicated to hide only a small set of nodes (community $C$ in the broad sense) targetly. In this paper, we address the problem of information over-mined in community detection from a global perspective. We aim at making the overall quantification of the communities detected in networks decrease significantly with subtle network changes, which shows the effectiveness of our attack strategies.

\section{Methods}
\label{section3}
First, we briefly demonstrate the major problem we focus on in this paper, using related notations and definitions. Let $G=(V,E)$ be the original network, where $V=\left\{v_1,v_2,...,v_n\right\}$ is the set of nodes and $E=\left\{e_1,e_2,...,e_m\right\}$ is the set of links. $E_{attack}=\left\{+\widetilde{e}_1,-\widetilde{e}_2,...,+\widetilde{e}_{2T-1},-\widetilde{e}_{2T}\right\}$ is the solution that comprised of a series of links, ``+" represents adding the link to the network while ``-" represents removing. Then, we have the adversarial network $G'=(V',E')$ defined as follows:
\begin{equation}
\label{eq:1}
\left\{
\begin{array}{lr}
V'=V & \\
E'=E+E_{attack}
\end{array}
\right.
\end{equation}
We thus want to find out such $E_{attack}$ to change some connections in $G$ and get the adversarial network $G'$. And for these $G'$s, community detection algorithms perform significantly worse, i.e., the quality of detection results decreases.

Fig.~\ref{fig:1} gives the overview of our work. In the following, we give the details of attack strategies for community detection algorithms proposed in this paper, including heuristics and evolutionary ones.

\subsection{Heuristic Attack Strategy}
\label{section3.1}
We have introduced that we take link attacks on the network to change some relations between nodes. In this paper, we choose \emph{rewiring} attack to maintain the degree of target nodes, i.e., adding a link to the target node while deleting one from it. Such rewiring attack is of relatively less cost, i.e., only two nodes need to change their degree under a rewiring attack, while four nodes may change their degree if we choose to remove a link between two nodes and add a link between another two at the same time. In social networks, one rewiring attack means getting a false friend while hiding a real one for the person chosen as the target node, the total number of friends of this person keeps unchanged, such attack way can maintain the person's status in the entire network, i.e., rewiring shows a good concealment of taking attacks on the network.

While thinking of attack strategies, we first come up with a Random Attack (RA) strategy, described in Algorithm~\ref{alg:1}, for comparison. Here, we randomly select $K$ nodes from $G$ to form a target node set $V_t$. We then randomly select a node $v_i\in{V_t}$ in each iteration and implement the following operation: deleting an existent link while adding nonexistent one for the target node. The neighbor set of the target node $v_i$ is denoted by $N_i=\left\{v_j|<v_i,v_j>\in E\right\}$, while its non-neighbor set is denoted by $\overline{N}_i$. The number of rewiring attacks (or the number of iterations), denoted by $T$, is another parameter that represents the attack cost. The procedure of random attack strategy is shown in following pseudo-code.

\begin{algorithm}[!t]
\caption{\textbf{Random Attack (RA)}}
\label{alg:1}
\KwIn{$G$,$K$,$T$}
\KwOut {$G'$}
$V_t \gets RandomK(G,K)$\;
$AttackNum=0$\;
\For {$AttackNum<T$}
{
$v_i \gets RandomOne(V_t)$\;
$N_i,\overline{N_i} \gets GetSet(G,v_i)$\;
\If{$N_i \neq \emptyset $ and $\overline{N_i} \neq \emptyset$}
{
$v_{del} \gets RandomOne(N_i)$\;
$v_{add} \gets RandomOne(\overline{N_i})$\;
remove $<v_i,v_{del}>$ from $E$ and add $<v_i,v_{add}>$ to $E$\;
Update $G$\;
$AttackNum=AttackNum+1$\;
}
}
Get the adversarial network $G'$ under $T$ rewiring attacks.
\end{algorithm}

A network with community structure often shows the characteristic of having a high density of intra-community links and a lower density of inter-community links. Deleting links within community and adding links between communities thus can weaken the community structure in the network. Therefore, having some knowledge of the community structure in advance can help to design more effective attack strategies. Based on this, we propose a heuristic attack strategy combined with the community division results via specific community detection algorithm, called Community Detection Attack (CDA), the procedure of which is shown in Algorithm~\ref{alg:2}.

\begin{algorithm}[!t]
\caption{\textbf{Community Detection Attack (CDA)}}
\label{alg:2}
\KwIn{$G$, $K$, $T$}
\KwOut {$G'$}
$C=\left\{C_1,C_2,...,C_h\right\} \gets Detection(G)$\;
$V_t \gets RandomK(G,K)$\;
$AttackNum=0$\;
\For {$AttackNum<T$}
{
$v_i \gets RandomOne(V_t)$\;
$S_1,S_2 \gets GetSet(G,v_i,C)$\;
\If{$S_1 \neq \emptyset $ and $S_2 \neq \emptyset$}
{
$v_{del} \gets RandomOne(S_1)$\;
$v_{add} \gets RandomOne(S_2)$\;
remove $<v_i,v_{del}>$ from $E$ and add $<v_i,v_{add}>$ to $E$\;
Update $G$\;
$AttackNum=AttackNum+1$\;
}
}
Get the adversarial network $G'$ under $T$ rewiring attacks.
\end{algorithm}

For target node $v_i$, we define its intra-community neighbor set as $S_1=C_i \cap N_i$, where $C_i$ is the node set of the community that $v_i$ belongs to. Similarly, we can get its inter-community non-neighbor set $S_2=\overline{C}_i \cap \overline{N}_i$, with $\overline{C}_i=V-C_i$.

\begin{figure}[!t]
\centering
\includegraphics[width=1\linewidth]{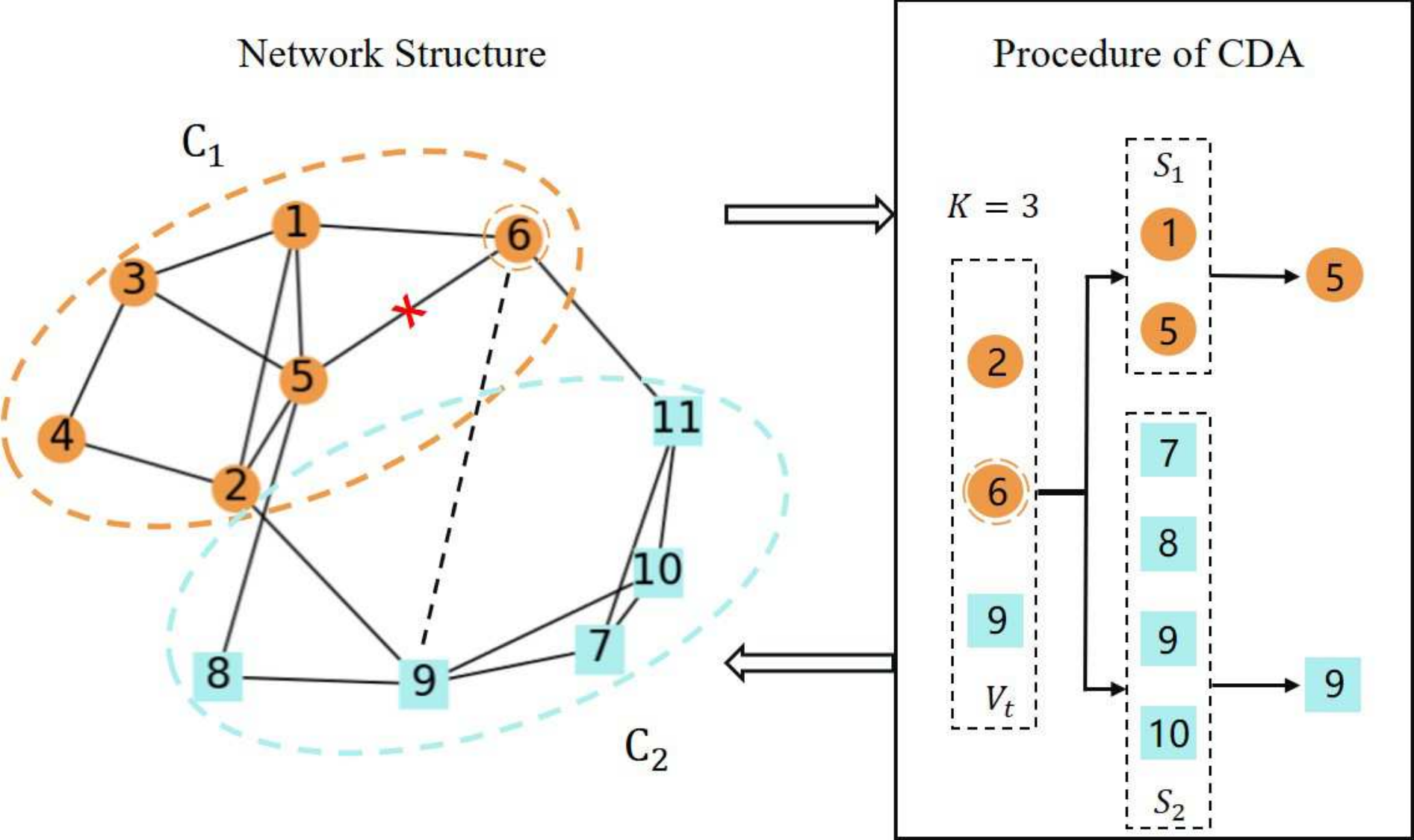}\caption{$C_1$ and $C_2$ represent the communities detected by a specific community detection algorithm and different colors of nodes represent different communities they belong to. Node 6 is chosen from the target node set $V_t$ as the target node under the current attack. $S_1$ and $S_2$ represent intra-community neighbor set and inter-community non-neighbor set of the target node, respectively. The line with a red ``$\times$'' represents the deleted  link and the dashed line represents the added link, performed by the attack.}
\label{fig:2}
\end{figure}

Fig.~\ref{fig:2} illustrates how the CDA strategy works graphically. For the target node, the removal of intra-community links weakens its connection with the original organization and the addition of inter-community links generates a force to drag it to the others.

It was found that many real-world networks followed power-law degree distribution~\cite{barabasi1999emergence}, consistent with the classic 80/20 rule. That means, usually, a small number of nodes possess a large number of connections, while the rest only have a few in real-world networks. Some changes of these hub nodes of larger degree usually have a bigger impact on the whole network structure. For instance, statistically, people with more friends in social networks always plays more important roles in the circle and their absence seems to make the party less active. Therefore, we propose another heuristic attack strategy, namely Degree Based Attack (DBA), aiming to attack the nodes of large degree in the network.

\begin{figure*}[!t]
\centering
\includegraphics[width=\linewidth]{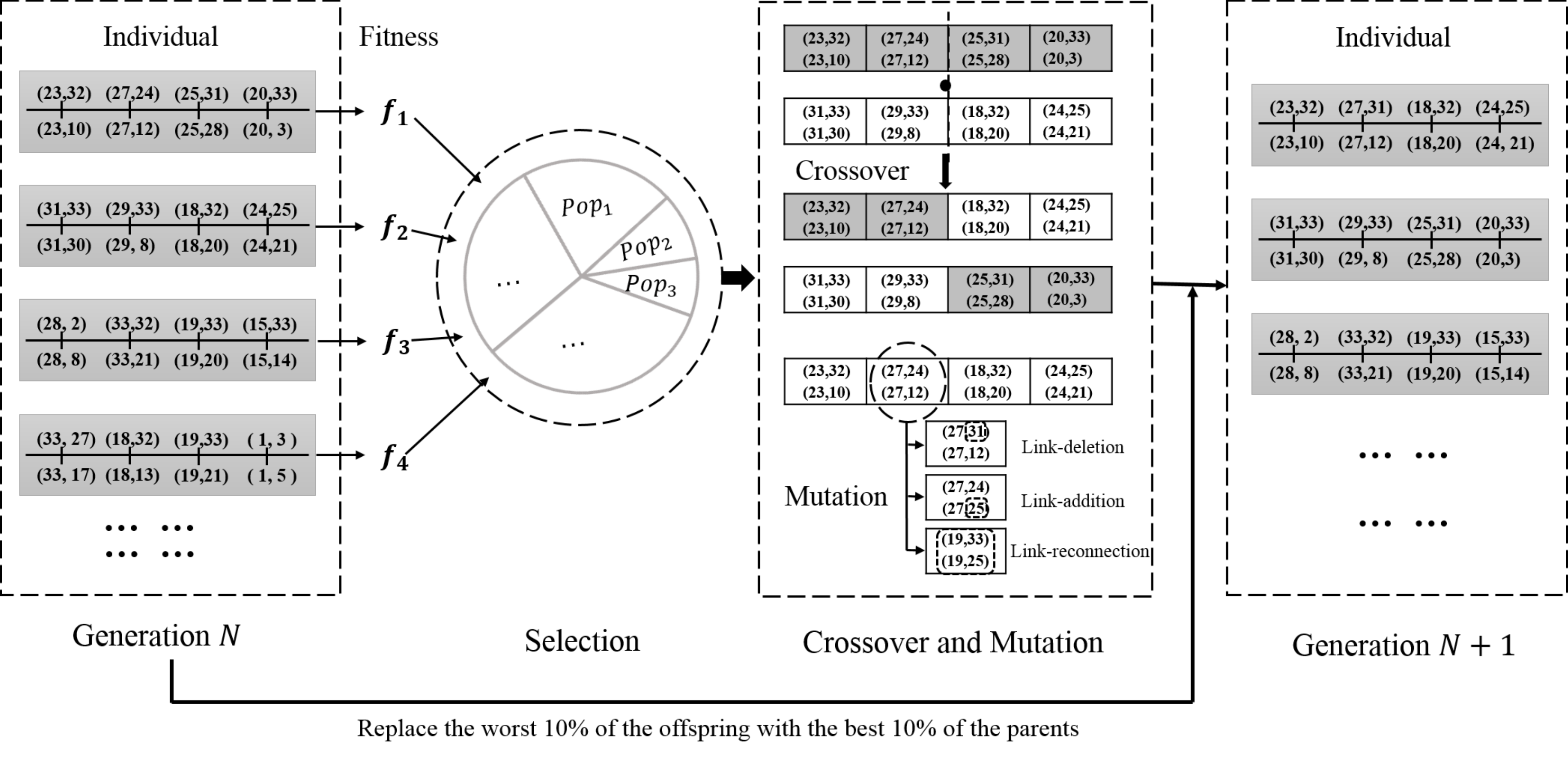}\caption{The overview of one iteration of GA based Q-Attack Strategy.}
\label{fig:3}
\end{figure*}

\begin{algorithm}[!t]
\caption{\textbf{Target Nodes Chosen Based on Degree}}
\label{alg:3}

$C=\left\{C_1,C_2,...,C_h\right\} \gets Detection(G)$\;
\For {$len(KNode)<K$}
{
$v \gets BiggestDegree(V)$\;
add node $v$ to $KNode$\;
$V=V-C_i$, where $v \in C_i$\;
\If{$K>h$}
{
update $V$ every turn\;
repeat from step $BiggestDegree$\;
}
}
\end{algorithm}

The only difference between DBA and CDA is that, in DBA we choose the nodes of large degree as our target nodes, while in CDA we do it randomly. Thus, next we just explain how to choose the $K$ target nodes in DBA based on node degree, since the rest attack part is totally the same as CDA. The procedure to choose the target nodes based on their degree in DBA is shown in Algorithm~\ref{alg:3}. We get the community detection results $C=\left\{C_1,C_2,...,C_h\right\}$ at the beginning of the algorithm. The function \emph{BiggestDegree} (line3) gets the node with the biggest degree from node set $V$ and the procedure in line 5 means removing the nodes belonging to $C_i$ from $V$. One turn is finished when we have iterated through all communities in $C$ and gotten the node with the biggest degree in each community. The step \emph{update $V$ every turn} (line 7) means removing the nodes that have been chosen as targets from the initial node set $V$, which consists of all the nodes in the network.

\subsection{GA Based Q-Attack}
\label{section3.2}
Evolutionary algorithms are a type of intelligent optimization technologies designed by simulating natural phenomena or leveraging various mechanisms of living organisms in nature~\cite{holland1975adaptation,kirkpatrick1983optimization,eberhart1995new}. GA is a typical one of such algorithms. Its basic model was firstly introduced and investigated by Holland in 1975~\cite{holland1975adaptation}, which mainly leveraged the natural selection mechanism to solve optimization problems. A population of chromosomes, which represent individuals with different characteristics, initialize at the beginning of the algorithm and then generate offspring according to the designed genetic operators, such as crossover and mutation. Individuals with better fitness have more chances to survive and multiply so that the population are able to evolve. Here we propose an attack strategy based on GA to search effective ¡°rewiring¡± attack schemes. Fig.~\ref{fig:3} gives an overview of the evolutionary process.

Before the implementation of GA, we first give our encoding method and fitness function.
\begin{itemize}
\item \textbf{Encoding:} Same as the attack strategies proposed in Sec.~\ref{section3.1}, here we choose a ¡°rewiring¡± attack as a gene, including a deleted link and an added link. The length of each chromosome represents the number of attacks.
\item \textbf{Fitness function:} Modularity is an important evaluation to assess the quality of a particular partition for the network and it has been widely applied in the field of community detection (more details in Sec.~\ref{section4.3}). Here we design our fitness function as follows:
\begin{equation}
\label{eq:2}
f=e^{-Q},
\end{equation}
indicating that individuals of lower modularity will have larger fitness. And this is also the reason why we name this attack strategy as \emph{Q-Attack}.
\end{itemize}

Now, we are going to find out individuals that can make the partition of an adversarial network under specific community detection algorithm have lower modularity. In the following, we will give a specific illustration about how GA operates in community detection attack.

\begin{itemize}
\item \textbf{Initialization.} An initial population is randomly generated at a fixed size in GA. One thing we need to take care is to avoid link deletion or addition conflict, which means deleting or adding a link repeatedly. Each individual in the population is a combination of \emph{rewiring} attacks, representing a solution of attacks on the network.
\item \textbf{Selection.} Selection is a manifestation of the \emph{survival of the fittest} mechanism. According to the laws of nature's evolution, the individuals of higher fitness have a greater chance to survive and multiply than those of lower fitness. We use a common selection operator, namely \emph{roulette wheel}. In this selection way, the probability for an individual to mating is proportional to its fitness, as represented by Eq.~(\ref{eq:3}):
\begin{equation}
\label{eq:3}
p_i=\frac{f(i)}{\sum_{j=1}^n f(j)}.
\end{equation}
\item \textbf{Crossover.} Here, we adopt single point crossover which is the easiest to achieve in GA, i.e., we generate a cut point at random and change the gene segments between parents. The probability for paired individuals to generate offspring through crossover operation is denoted by $P_c$. In order to avoid the conflicts on genes (as mentioned before, two identical deleted-links or added-links are not allowed to appear on one chromosome), we do a feasibility test before two individuals get crossed.
\item \textbf{Mutation.} To prevent the solution from falling into a local optimum, mutation operator is a necessary component in GA. Considering the network characteristics, here, we propose three kinds of mutation operators: link-deletion mutation, link-addition mutation and link-reconnection mutation. link-deletion or link-addition mutation indicates that the deleted or added link of the gene changes independently, while link-reconnection mutation indicates that a gene is completely replaced. These three mutations occur with equal probability and the overall mutation rate is denoted by $P_m$. Such designs on mutation operators can promote the diversity of population. Conflicts detection is also considered while the genes are mutating.
\item \textbf{Elitism.} Over the course of evolution, individuals with high fitness in the parent may be destroyed. Elitism strategy has been proposed to solve this problem. In this paper, we replace the worst 10\% of the offspring with the best 10\% of the parent to maintain excellent genes.
\item \textbf{Termination criteria.} We set the evolutionary generation as a constant and the algorithm stops when this condition is fulfilled.
\end{itemize}

\begin{algorithm}[!t]
\caption{\textbf{Q-Attack Based on GA}}
\label{alg:4}
\KwIn{$G$,$PopSize$,$Generation$,$P_c$,$P_m$,$T$}
\KwOut {$Pop$}
$ParentPop \gets \textbf{Inatialization}(G,PopSize,T)$\;
\While{$i<Generation$}
{
$SelectedPop \leftarrow \mathbf{Selection}(ParentPop,PopSize)$\;
$CrossoverPop \leftarrow \mathbf{Crossover}(SelectedPop,PopSize,T,P_c)$\;
$MutationPop \leftarrow \textbf{Mutation}(G,CrossoverPop,PopSize,T,P_m)$\;
$OffspringPop \leftarrow \textbf{Elistism}(MutationPop,ParentPop)$\;
$ParentPop \leftarrow OffspringPop$\;
$i \leftarrow i+1$\;
}
\end{algorithm}

In summary, GA based Q-Attack mainly involves the encoding, fitness function and the design of genetic operators. The pseudo-code of this attack strategy is shown in Algorithm~\ref{alg:4}.

\section{Experiments}
\label{section4}
\subsection{Datasets}
\label{section4.1}
In order to evaluate the attack effect of our attack strategies, we test them on four real-world datasets, including Zachary's karate club network~\cite{zachary1977information}, bottlenose dolphin network~\cite{lusseau2003bottlenose}, American college football network~\cite{girvan2002community} and American political books network~\cite{newman2006modularity}, which are commonly used for community detection with ground-truth partitions.
\begin{itemize}
\item \textbf{Zachary's karate club network:} It is a dataset about the relationship in a karate club recorded by Zachary. The network consists of 34 nodes and 78 links, which is ultimately divided into two groups out of the conflicts between the instructor and the administrator.
\item \textbf{Bottlenose dolphin network:} This network is created by Lusseau, demonstrating the associations of the dolphins living off Doubtful Sound, New Zealand. The network consists of 62 nodes and 159 links. In reality, these dolphins are divided into two groups.
\item \textbf{American college football network:} This dataset is about the college football matches in one season. There are 115 nodes and 613 links in the network. The nodes which represent the college football teams belong to 12 communities.
\item \textbf{American political books network:} The network consists of 105 nodes and 441 links. Each node in the network represents a book published on political topics and the link between two books shows that they are purchased by the same consumer. These books are divided into three categories according to their political stands.
\end{itemize}

\subsection{Community Detection Algorithms}
\label{section4.2}
Here, we introduce six well-known community detection algorithms that are attacked by our strategies proposed in Sec.~\ref{section3}. Note that these algorithms are mainly used for detecting non-overlapping community structures.
\begin{itemize}
\item \textbf{Fast Newman Algorithm (FN)~\cite{newman2004fast}:} It is a hierarchical agglomerative algorithm. Each node is considered as a community at the beginning and we merge communities in pairs repeatedly with the aim of optimizing $Q$, until all the nodes are merged into one community.
\item \textbf{Spectral Optimization Algorithm (SOA)~\cite{newman2006modularity,newman2006finding}:} Through analyzing the spectral properties of networks, Newman gives a reformulation of modularity:
\begin{equation}
\label{eq:4}
Q=\frac{1}{4m}\bm{s}^T(A_{i j}-\frac{k_i k_j}{2m})\bm{s}=\frac{1}{4m}\bm{s}^T\bm{B}\bm{s}
\end{equation}
where $m$ is the number of links, $A_{i j}$ is the element of adjacency matrix $A$, $\bm{s}$ is a column vector with each element $s_i$ representing the community label of node $v_i$, $k_i$ and $k_j$ are the degrees of nodes $v_i$ and $v_j$, respectively. $\bm{B}$ is called modularity matrix. The communities that nodes belong to depend on the signs of the elements in the leading eigenvector of the modularity matrix.
\item \textbf{Louvain Algorithm (LOU)~\cite{blondel2008fast}:} This algorithm considers the community detection as a multi-level modularity optimization problem. It starts with isolated nodes and repeats the process of removing the node to the community which obtains maximum gain of modularity until no further improvement can be achieved. Then it constructs a new network by merging the community into a super-node. These two steps are performed repeatedly until the algorithm is stable.
\item \textbf{Label Propagation Algorithm (LPA)~\cite{raghavan2007near}:} Each node is assigned a label at the beginning of the algorithm and it is updated as the most frequent label of the node's neighbors during the information propagation process. The algorithm terminates when no node changes its label anymore.
\item \textbf{Infomap Algorithm (INF)~\cite{rosvall2008maps}:} This algorithm is based on network maps and coding theory. It detects communities by compressing the description of information flows on networks. In other words, the key of Infomap is to minimize the average description length $L(M)$ since more information is generated when random walks between communities occur.
\item \textbf{Node2vec+KMeans (Node2vec+KM)~\cite{grover2016node2vec}:} Node2vec is one of the state-of-the-art embedding methods and it learns low-dimensional representation for nodes in the network. K-means clustering algorithm is then used to detect communities when nodes are embedded into Euclidean space through node2vec.
\end{itemize}

\subsection{Metrics for Community Structure}
\label{section4.3}
\begin{itemize}
\item \textbf{Modularity:} It is widely used to measure the quality of divisions of a network, especially for the networks with unknown community structure, which was firstly proposed by Newman and Grivan~\cite{newman2004finding} and is defined by
\begin{equation}
\label{eq:5}
Q=\sum_i(e_{ii}-a_i^2),
\end{equation}
where $e_{ii}$ represents the fraction of links in the network with two terminals both in cluster $C_i$, $a_i=\sum_je_{ij}$ represents the fraction of links that connect to the nodes in cluster $C_i$. In other words, modularity $Q$ measures the difference between actual fraction of within-community links and the expected value of the same quantity with random connections. A reformulation of the modularity is represented by Eq.~(\ref{eq:4}).
\item \textbf{Normalized Mutual Information (NMI):} It is another commonly used criterion to assess the quality of clustering results in analyzing network community structure, which was proposed by Danon et al.~\cite{danon2005comparing}. For two partitions $X$ and $Y$, the mutual information $I(X,Y)$ is defined as the relative entropy between the joint distribution $P(X£¬Y)$ and the production distribution $P(X)P(Y)$:
\begin{eqnarray}
\label{eq:7}
I(X,Y) &=&D(P(X,Y)||P(X)P(Y)) \nonumber \\
&=&\sum_{x,y}p(x,y)\log\frac{p(x,y)}{p(x)p(y)}.
\end{eqnarray}
A noticeable problem for mutual information alone being a similarity measure is that subpartitions derived from $Y$ by splitting some of its clusters into small ones would have same mutual information with $Y$. NMI thus is proposed to deal with this problem, which is defined by
\begin{equation}
\label{eq:8}
I_{norm}(X,Y)=\frac{2I(X,Y)}{H(X)+H(Y)}.
\end{equation}
The value of NMI indicates the similarity between two partitions, i.e., the larger value means  more similar between the two. When NMI equals 1, partition $X$ is identical to partition $Y$.
\end{itemize}

\begin{figure}[!t]
\centering
\includegraphics[width=\linewidth]{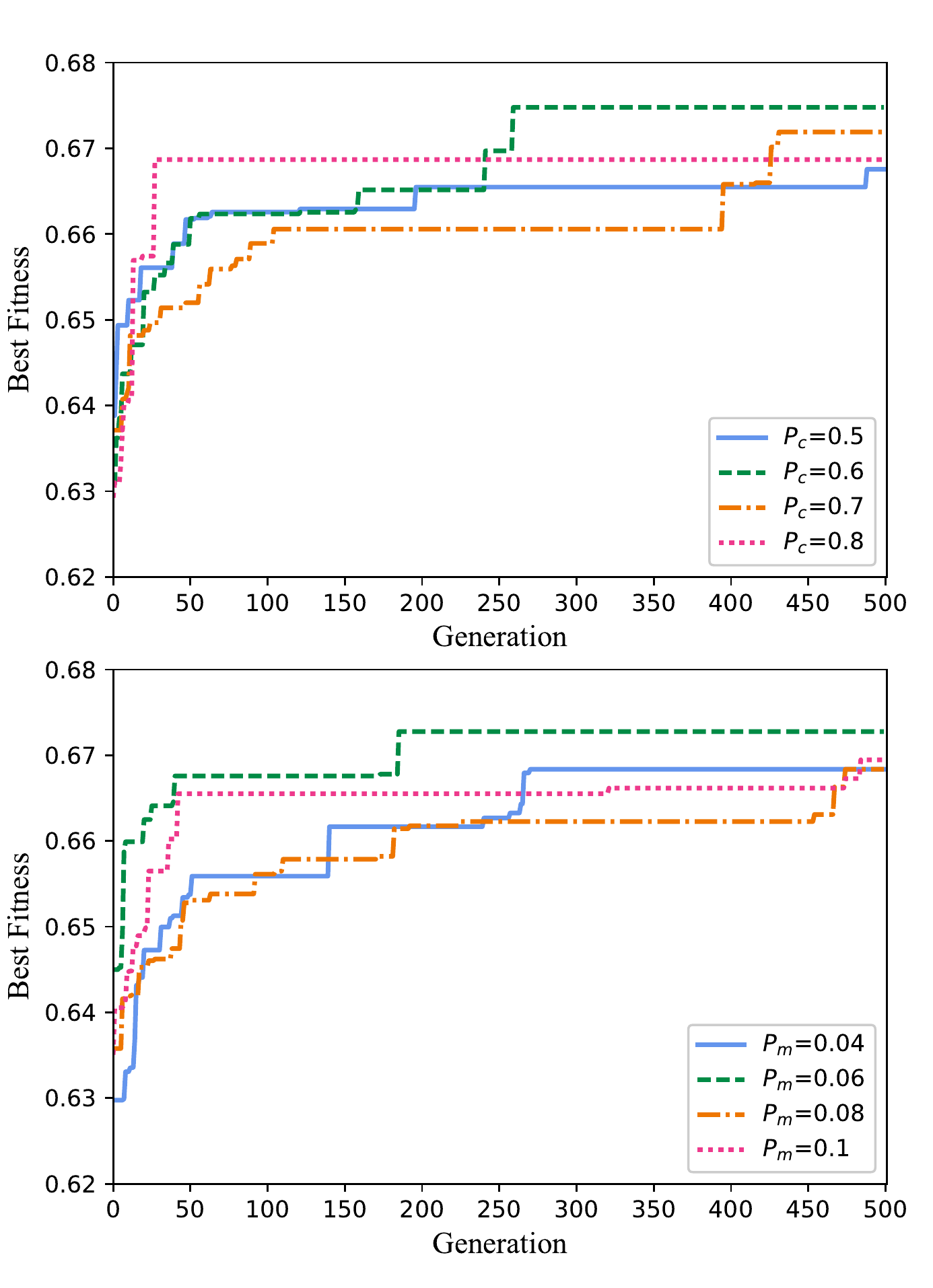}\caption{The curves of best fitness for various values of parameters $P_c$ and $P_m$ when Q-Attack is applied to FN algorithm on Dolphins network.}
\label{fig:6}
\end{figure}

\begin{table}[!t]
\renewcommand{\arraystretch}{1.2}
\setlength{\abovecaptionskip}{0pt}
\setlength{\belowcaptionskip}{5pt}
\centering
\caption{Optimal parameters for Q-Attack in different cases}
\setlength{\tabcolsep}{1.3mm}{
\begin{tabular}{ccccccc} \hline\hline
\label{table1}
 & FN+Kar & FN+Dol  & SOA+Kar  & SOA+Dol & LOU+Kar & LOU+Dol \\\hline
$P_c$	& 0.8 & 0.6 &	0.7 & 0.6 & 0.7 & 0.8\\
$P_m$	& 0.1 & 0.06 &	0.1 & 0.04 & 0.1 & 0.1\\\hline\hline
\end{tabular}}
\end{table}

\begin{figure*}[!t]
\centering
\includegraphics[width=\linewidth]{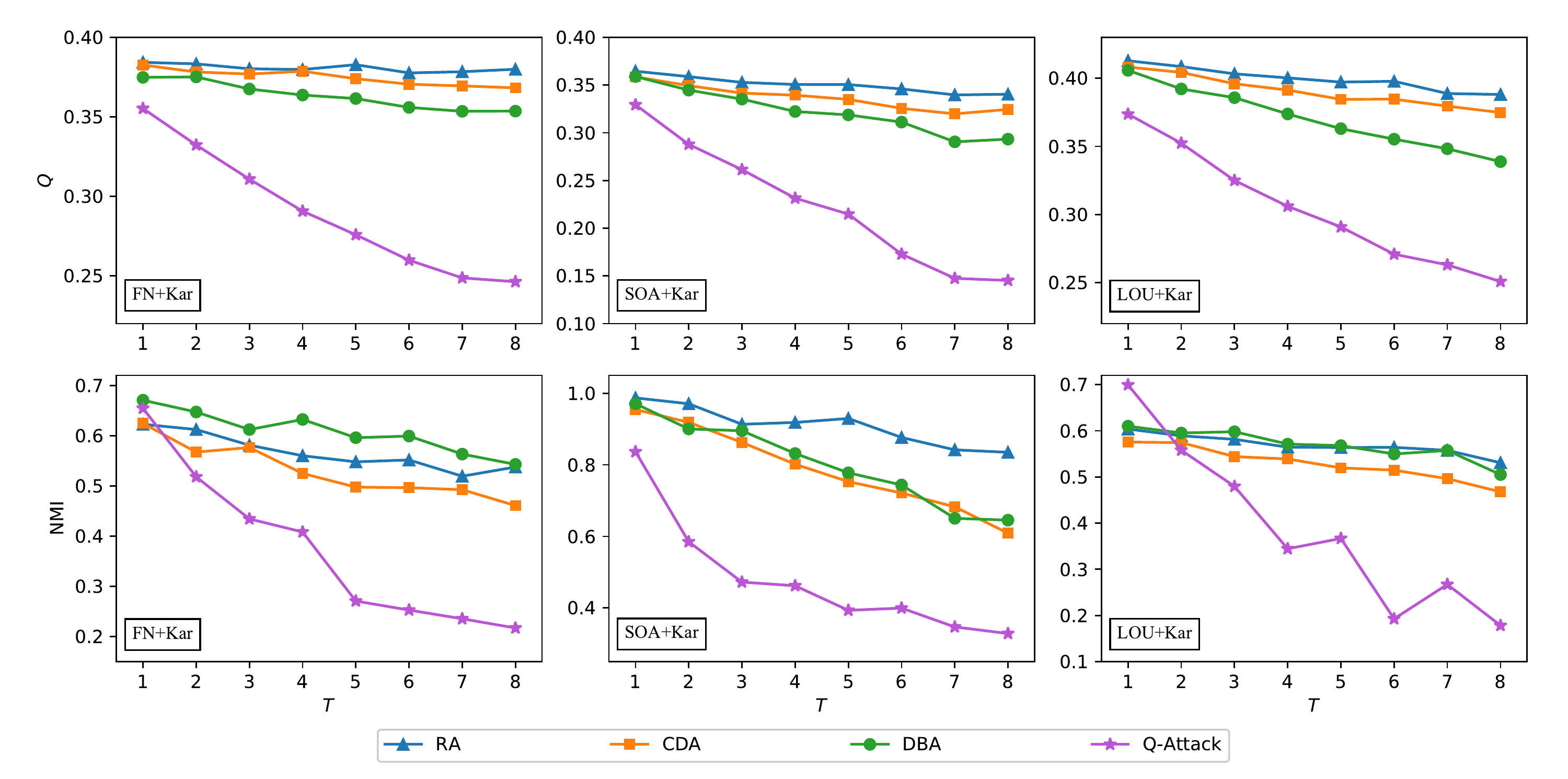}\caption{The attack effects of the four attack strategies on three community detection algorithms and Karate network, for various numbers of attacks.}
\label{fig:7}
\end{figure*}

\begin{figure*}[!t]
\centering
\includegraphics[width=\linewidth]{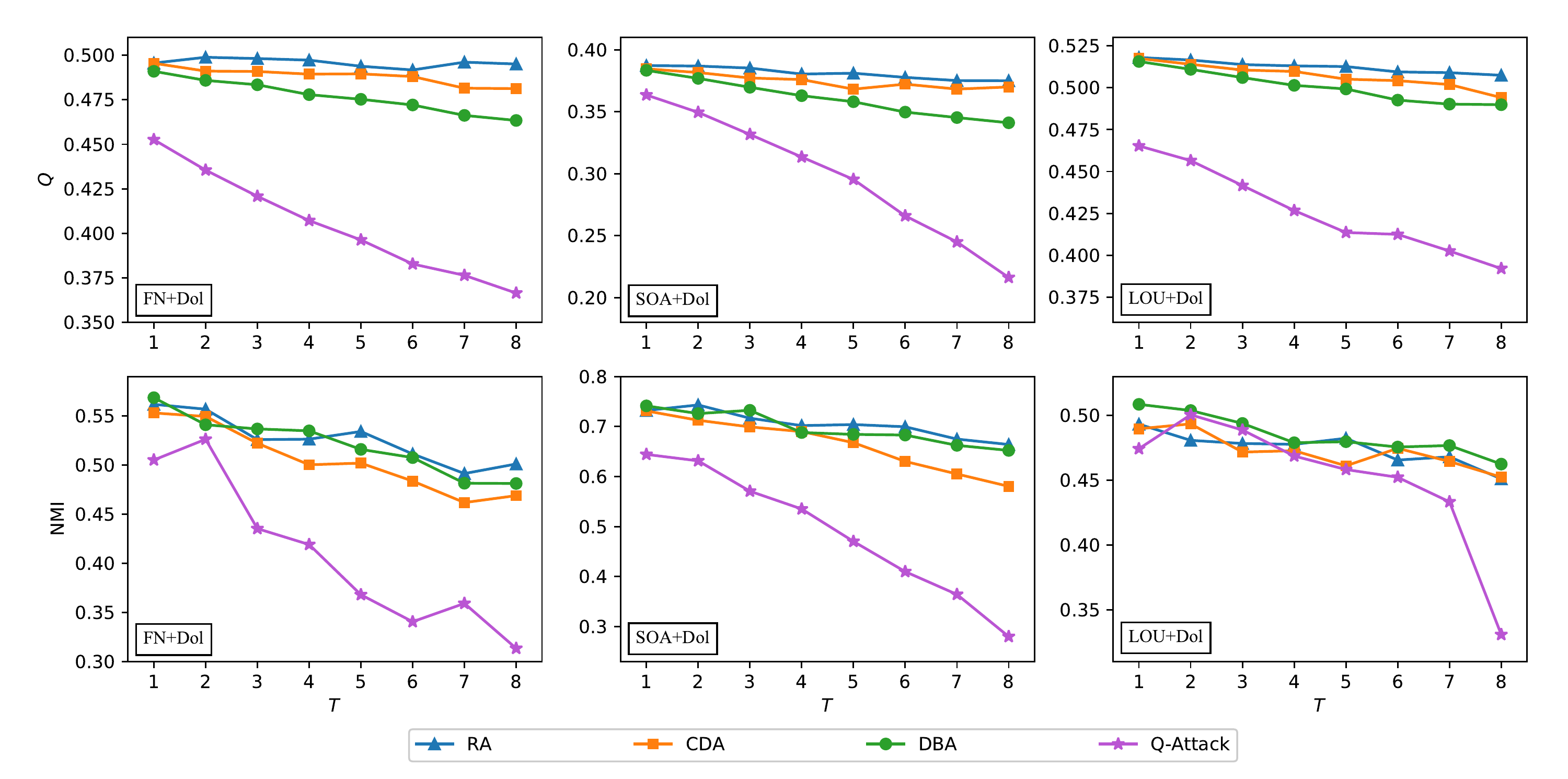}\caption{The attack effects of the four attack strategies on three community detection algorithms and Dolphins network, for various numbers of attacks.}
\label{fig:10}
\end{figure*}

\subsection{Experimental Results}
\label{section4.4}
Our experiments are performed on a PC i5 CPU 2.60GHz and 4GBs RAM. Our programming environment is python 2.7. We firstly test our attack strategies on three community detection algorithms and two networks with different attack number T. For our GA based Q-Attack strategy, it concludes four parameters. We set the population size $PopSize=100$ and the evolutionary generation $Generation=500$ uniformly. Then, we tune the crossover rate $P_c$ and the mutation rate $P_m$ from four different values (0.5, 0.6, 0.7, 0.8 for $P_c$ and 0.04, 0.06, 0.08, 0.1 for $P_m$), respectively, according to the results for parameter configuration. Fig.~\ref{fig:6} shows the curves of best fitness for a set of experiments when Q-Attack is applied to FN algorithm on Dolphins, and we get the optimal parameters $P_c=0.6$ and $P_m=0.06$ for this case, under $T=4$, which is a medium value. Table~\ref{table1} gives all the optimal parameters we use for Q-Attack in six different situations here. And for heuristic attack strategies proposed in Sec.~\ref{section3.1}, the numbers of target nodes for Karate and Dolphins are set to 5 (15\% nodes) and 6 (10\% nodes) respectively.

\begin{table*}[htbp]\small
\renewcommand{\arraystretch}{1.3}
\setlength{\abovecaptionskip}{0pt}
\setlength{\belowcaptionskip}{5pt}
\centering
\caption{Attack results when the attack number is set to a fixed percentage of links in the networks. Here, we only present the relative reduction of $Q$ and NMI, compared with their values in the original network without attack.}
\setlength{\tabcolsep}{0.2mm}{
\begin{tabular}{c|ccccccc|ccccccc} \hline\hline
\label{table2}
\multirow{2}{*}{Karate}&\multicolumn{7}{c|}{$Q$}&\multicolumn{7}{c}{NMI}\\\cline{2-15}
& FN & SOA & LOU & LPA & INF & Node2vec+KM & Average & FN & SOA & LOU & LPA & INF & Node2vec+KM & Average\\\hline
 Original& 0.381 & 0.371  & 0.419 & 0.380 & 0.402 & 0.368 & 0.387&0.692 &1  &	0.587 	&0.806 &0.699 &0.951&0.789  \\
 RA & 0.24\% & 5.62\%  & 4.43\% & 18.13\% & 3.64\% & 6.47\% & 6.42\%&19.14\%&8.16\%&3.79\%&	27.58\%&	9.43\%&	13.30\%&	13.57\% \\
 CDA &0.53\%&8.64\%&6.56\%&23.89\%&12.93\%	&9.78\%	&10.39\% &\textbf{24.23\%} &\textbf{19.84\%}	&\textbf{8.14\%}	&\textbf{34.45\%}	&\textbf{23.76\%}&	\textbf{23.37\%}	&\textbf{22.30\%} \\
DBA &\textbf{4.46\%}& \textbf{13.24\%}&\textbf{10.75\%}&\textbf{26.48\%}&\textbf{15.93\%}&\textbf{14.53\%}&\textbf{14.23\%}&8.71\%&16.84\%&	2.65\%	&33.12\%&	7.33\%&	19.22\%	&14.65\% \\
Q-Attack&\textbf{23.64\%}&\textbf{37.70\%}&	\textbf{26.92\%}&\textbf{36.95\%}	&\textbf{74.28\%}	&\textbf{43.74\%}	&\textbf{40.54\%}&	\textbf{41.06\%}&\textbf{53.81\%}	&\textbf{41.30\%}&	\textbf{35.54\%}&	\textbf{68.07\%}	&\textbf{63.92\%}&	\textbf{50.62\%} \\
\hline
\multirow{2}{*}{Dolphins}&\multicolumn{7}{c|}{$Q$}&\multicolumn{7}{c}{NMI}\\\cline{2-15}
& FN & SOA & LOU & LPA & INF & Node2vec+KM & Average & FN & SOA & LOU & LPA & INF & Node2vec+KM & Average\\\hline
Original&0.495 &0.390&	0.519 &	0.506 &	0.524 &	0.379 &	0.469 &	0.573&0.753  	&0.511 &	0.607& 	0.553 &	0.889 &	0.648 \\
RA	&0.09\%&3.81\%&	2.15\%&	9.83\%&	3.19\%&	3.23\%&	3.72\% &12.52\%&11.84\%		&\textbf{11.71\%}&	\textbf{5.28\%}&	13.47\%	&9.22\%&	10.67\%\\
CDA&	2.88\%&	5.10\%	&4.71\%	&17.74\%&	6.76\%	&4.20\%&	6.90\%&	\textbf{18.14\%}&\textbf{22.94\%}&	11.45\%	&1.47\%	&\textbf{16.53\%}&	\textbf{19.41\%}	&\textbf{14.99\%} \\
DBA&\textbf{6.49\%}&\textbf{12.51\%}&\textbf{5.53\%}	&\textbf{18.26\%}&	\textbf{8.72\%}	&\textbf{11.52\%}&	\textbf{10.50\%}&15.98\%&13.41\%	&9.48\%	&2.41\%&	15.80\%&	14.35\%	&11.91\%\\
Q-Attack&\textbf{26.04\%}&	\textbf{44.53\%}	&	\textbf{24.38\%}	&\textbf{22.63\%}&	\textbf{12.84\%}	&\textbf{19.59\%}&	\textbf{25.00\%} &	\textbf{45.27\%}&\textbf{62.83\%}&	\textbf{35.24\%}	&\textbf{6.66\%}	&\textbf{15.97\%}&	\textbf{39.93\%}&	\textbf{34.32\%}\\
\hline\hline
\multirow{2}{*}{Football}&\multicolumn{7}{c|}{$Q$}&\multicolumn{7}{c}{NMI}\\\cline{2-15}
& FN & SOA & LOU & LPA & INF & Node2vec+KM & Average & FN & SOA & LOU & LPA & INF & Node2vec+KM & Average\\\hline
Original & 0.550 &0.493 &0.605 &0.603 &0.601 &0.598 &0.575&	0.698&0.699  &	0.890 	&0.888 &	0.924 	&0.922& 	0.837  \\
RA&-0.05\%	&-3.26\%&1.89\%&4.59\%&1.94\%&	1.91\%&	1.17\% &-4.26\%&-5.38\%	&	1.94\%	&1.08\%	&0.00\%	&0.05\%	&-1.09\%\\
CDA&\textbf{1.86\%}&	-0.65\%&	2.97\%&	5.31\%&	3.16\%	&3.25\%	&\textbf{2.65\%}&-1.18\%&\textbf{1.63\%}	&	\textbf{2.64\%}	&\textbf{1.60\%}&	\textbf{0.12\%}&	0.24\%&	\textbf{0.84\%} \\
DBA&0.56\%&	\textbf{-0.03\%}		&\textbf{3.16\%}&	\textbf{5.47\%}&	\textbf{3.22\%}&	\textbf{3.28\%}	&2.61\%&	\textbf{0.13\%}&-1.49\%	&2.44\%	&1.49\%	&\textbf{0.11\%}	&\textbf{0.39\%}&	0.51\% \\
Q-Attack&\textbf{20.36\%}&	\textbf{19.97\%}	&\textbf{11.02\%}	&\textbf{9.12\%}&	\textbf{3.43\%}&	\textbf{4.59\%}&	\textbf{11.42\%}&	\textbf{27.10\%}&\textbf{34.23\%}&	\textbf{16.01\%}	&\textbf{6.93\%}&	-0.20\%	&\textbf{2.68\%}	&\textbf{14.46\%} \\
\hline
\multirow{2}{*}{Polbooks}&\multicolumn{7}{c|}{$Q$}&\multicolumn{7}{c}{NMI}\\\cline{2-15}
& FN & SOA & LOU & LPA & INF & Node2vec+KM & Average & FN & SOA & LOU & LPA & INF & Node2vec+KM & Average\\\hline
Original& 0.502&0.467 &	0.520 &	0.495 	&0.523 &	0.500 &	0.501&0.531&0.520  &	0.512 &	0.586 &	0.493 &	0.564 &	0.534  \\
RA&1.26\%&1.66\%		&1.76\%&	2.01\%&	1.59\%&	1.62\%&	1.65\%&0.11\%&0.71\%&	1.60\%	&4.74\%	&0.10\%&	1.41\%&	1.45\% \\
CDA&2.95\%&	\textbf{3.96\%}&	2.75\%&	3.89\%&	2.64\%&	3.05\%&	3.21\%&	\textbf{2.89\%}&1.01\%&	\textbf{3.50\%}	&7.94\%	&\textbf{1.07\%}	&\textbf{5.00\%}	&\textbf{3.57\%}\\
DBA&\textbf{3.80\%}&2.44\%	&\textbf{3.02\%}&	\textbf{5.30\%}&	\textbf{3.29\%}	&\textbf{3.31\%}&	\textbf{3.53\%}&0.72\%&\textbf{2.50\%}	&-1.17\%&\textbf{9.10\%}	&-1.50\%&2.20\%	&1.98\%\\
Q-Attack&\textbf{22.17\%} &	\textbf{16.35\%}&\textbf{17.92\%}	&\textbf{5.40\%}&	\textbf{7.48\%}&	\textbf{5.82\%}	&\textbf{12.52\%}&	\textbf{40.53\%}&\textbf{21.29\%}	&\textbf{37.06\%}&	\textbf{13.02\%}&	\textbf{0.56\%}&	\textbf{7.97\%}&	\textbf{20.07\%} \\
\hline\hline
\end{tabular}}
\end{table*}

Modularity $Q$ and NMI introduced in Sec.~\ref{section4.3} are taken as the metrics to evaluate effectiveness of attack strategies. The results are shown in Figs.~\ref{fig:7} and \ref{fig:10}. Corresponding to each specified $T$, we generate 50 adversarial networks for each heuristic strategy and record the mean values of $Q$ and NMI. Each point on the curve of Q-Attack is the mean value over 10 runnings with $T$ changing from 2 to 8. While for $T=1$,  crossover operator described in GA is hard to implement. Here are two solutions for this problem: 1) we can perform an internal crossover for one \emph{rewiring} attack, i.e. swapping out the deleted link or the added one; 2) we can get the best result by going through all possible situations. Here, we get the results according to the second solution, since the two networks here we use are both small and thus it only takes little time to get the results.

\begin{figure*}[!t]
\centering
\includegraphics[width=.9\linewidth]{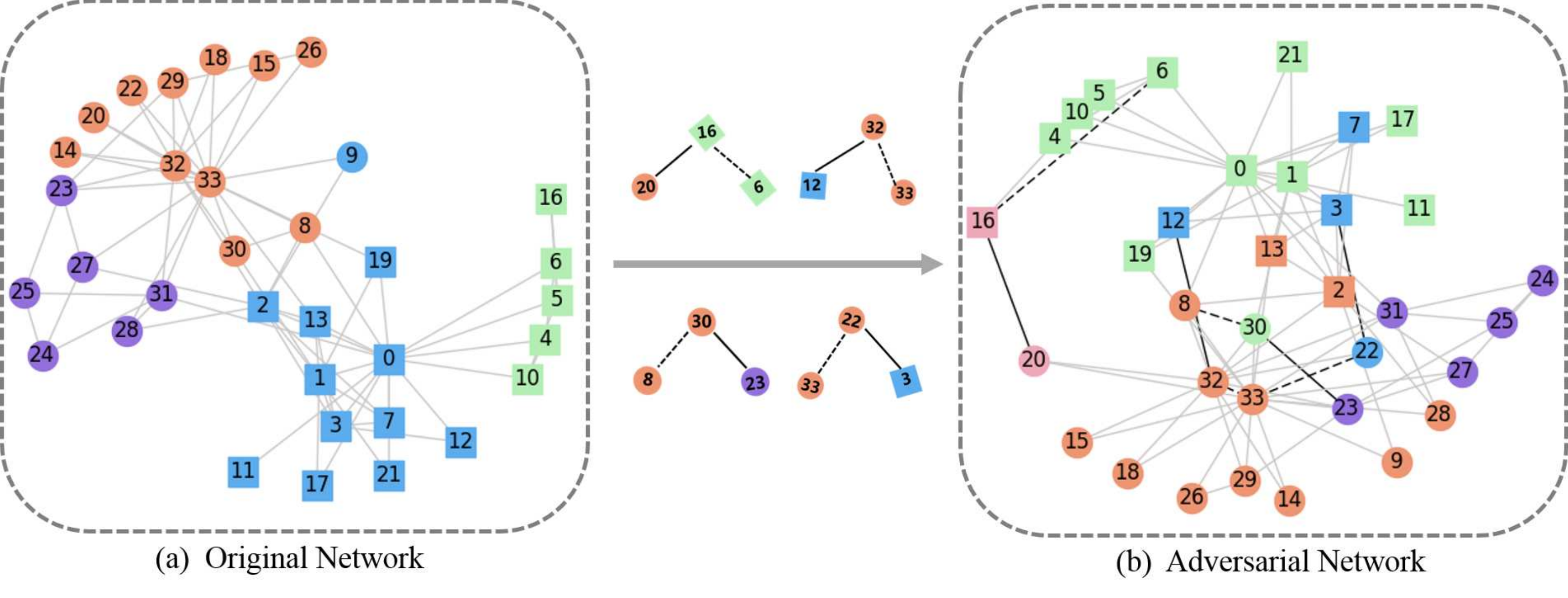}\caption{(a) Communities detected by Louvin algorithm on Karate network.(b) Communities detected by Louvin algorithm on the adversarial network. Different colors of nodes represent the different communities they belong to, while different shapes represent the different communities these nodes belong to according to the given label.}
\label{fig:8}
\end{figure*}

From Figs.~\ref{fig:7} and \ref{fig:10}, we can see that, in general, for almost all attack strategies, the values of $Q$ and NMI decrease as the budget $T$ increases, i.e., the attack effect is more significant when more links can be changed. As expected, Q-Attack behaves best to reduce the value of $Q$. This is not surprising since its goal is designed to minimize $Q$, based on Eq.~(\ref{eq:2}). Moreover, for the other metric NMI, we can still find that the curves for Q-Attack present sharper decline than the others, indicating its best attack effect on both Karate and Dolphins networks. Such results validate that our GA based Q-Attack is indeed more effective in both weakening the strength of community structure and reducing the similarity between the community detection results and the true labels. Besides, we also notice that, in most situations, DBA does a better job than CDA in reducing $Q$ while we get the opposite results on the metric NMI. This seems reasonable by considering that nodes of large degree plays a vital role in maintaining the structure of network and thus attacks on such nodes may affect the network structure more significantly. However, on the other hand, such large-degree nodes always stay in the centers of communities, and thus are relatively difficult to be  grouped into wrong communities when they are under attack, which may lead to the less reduction of NMI for DBA than CDA.

To provide more comprehensive and quantitative comparisons, we take extensive experiments on six community detection algorithms and four networks totally. Crossover rate $P_c$ and mutation rate $P_m$ are set to 0.8 and 0.1, respectively. The parameter $K$, i.e. the number of nodes chosen as targets in heuristic attack strategies, is set to 11 (10\% nodes) for Football and 10 (10\% nodes) for Polbooks. The attack number $T$ is set to 5\% of total number of links in the network for Karate and Dolphins, 2\% for larger networks Football and Polbooks. We use the default settings for the parameters in node2vec and take the number of clusters in K-means the same as ground truth. Table~\ref{table2} shows the relative reductions of $Q$ and NMI under four attack strategies. For each column in the table, we mark the top two best results as bold. Again, we see that our GA based Q-Attack has significantly better attack effects, in terms of larger relative reductions of both $Q$ and NMI, than the three heuristic strategies in almost all cases except the reduction of NMI for Football while attacking on Infomap algorithm. We also observe that the advantage of GA based Q-Attack seems more obvious when we control the attack budget to a smaller value. The relative reduction of two metrics for Q-Attack is about 2 to 3 times those for the optimal heuristic attack strategy on average with attack number $T$ limited to 5\% for Karate and Dolphins while it increases to 5 times and even more when attack number $T$ is limited to 2\% for Football and Polbooks.

In Fig.\ref{fig:8}, we visualize an example that Q-Attack is applied on the Louvin algorithm and Karate network. The community structure in the original network detected by Louvin algorithm is shown in Fig.~\ref{fig:8} (a). The modularity $Q$ and NMI are 0.4188 and 0.5866, respectively. We also use Louvin algorithm to detect the community structure in the adversarial network obtained under 4 rewiring attacks, which is shown in Fig.~\ref{fig:8} (b). After attack, the modularity $Q$ falls to 0.3053 while NMI falls to 0.3078 and the number of communities changes from 4 to 5. We can see that, indeed, the community detection result presents higher-level disorder after attack, indicating the good attack effect of Q-Attack.

\begin{table*}[!t]\renewcommand{\arraystretch}{1.3}
\small
\setlength{\abovecaptionskip}{0pt}
\setlength{\belowcaptionskip}{3pt}
\centering
\caption{Analysis on the transferability of Q-Attack. Here, we use the adversarial networks, generated by Q-Attack on particular community detection algorithms, to attack other algorithms.}
\label{tab:trans}
\setlength{\tabcolsep}{1.5mm}{
\begin{tabular}{c|c|cccccccc} \hline\hline
Dataset&Network&FN&SOA&LOU&LPA&INF&Node2vec+KM&Average&Average(-self) \\
\hline
\multirow{6}{*}{Karate}&Q-Attack(FN)&\textbf{23.64\%}&\textbf{13.84\%}&10.29\%&36.91\%&20.88\%&	18.33\%	&20.65\%&\textbf{20.05\%}\\
&Q-Attack(SOA)	&1.00\%	&\textbf{37.70\%}&4.52\%&27.98\%&14.49\% &\textbf{25.00\%}&18.45\%&14.60\% \\
&Q-Attack(LOU)&	10.29\%&	12.72\%&\textbf{26.92\%}&31.57\%&\textbf{34.67\%}	&15.14\%	&\textbf{21.89\%}	&\textbf{20.88\%}\\
&Q-Attack(LPA)&	1.19\%&	9.52\%&	7.59\%&	36.95\%&	6.11\%&	12.76\%	&12.35\%	&7.43\%\\
&Q-Attack(INF)&	\textbf{10.78\%}&	13.65\%&\textbf{15.23\%}&\textbf{43.69\%}&	\textbf{74.28\%}	&13.68\%	&\textbf{28.55\%}	&19.41\%\\
&Q-Attack(Node2vec+KM)&	0.35\%&	13.27\%&	6.04\%&\textbf{40.50\%}&	11.14\%	&\textbf{43.74\%}	&19.17\%&14.26\% \\
\hline
\multirow{6}{*}{Dolphins}&Q-Attack(FN)&	\textbf{26.04\%}&10.55\%&7.32\%&\textbf{25.17\%}&	7.48\%&9.48\%	&\textbf{14.34\%}	&\textbf{12.00\%}\\
&Q-Attack(SOA)&	3.92\%&\textbf{44.43\%}&5.12\%&14.54\%&6.25\%	&\textbf{24.35\%}&\textbf{16.43\%}&10.83\%\\
&Q-Attack(LOU)&	8.08\%&	10.85\%	&\textbf{24.96\%}&19.82\%&	\textbf{10.24\%}	&9.89\%	&13.97\%&	11.78\%\\
&Q-Attack(LPA)&	5.61\%	&7.73\%	&7.06\%	&\textbf{22.63\%}&9.15\%	&7.27\%	&9.91\%	&7.36\% \\
&Q-Attack(INF)&	\textbf{9.01\%}&9.83\%&\textbf{9.42\%}&\textbf{22.63\%}&\textbf{12.84\%}&9.50\%&	12.20\%&\textbf{12.08\%} \\
&Q-Attack(Node2vec+KM)&	-0.49\%	&\textbf{11.45\%}&1.53\%&21.90\%	&3.12\%	&\textbf{19.59\%}&9.52\%&	7.50\%\\
\hline
\multirow{6}{*}{Football}&Q-Attack(FN)&\textbf{20.36\%}&-1.78\%&\textbf{2.85\%}&5.58\%&2.87\%	&3.19\%	&\textbf{5.51\%}	&2.54\%\\
&Q-Attack(SOA)&	0.44\%&\textbf{19.97\%}&2.81\%&	\textbf{5.83\%}&	2.59\%&	3.06\%&	\textbf{5.78\%}&	\textbf{2.95\%}\\
&Q-Attack(LOU)&	\textbf{3.13\%}&\textbf{-0.34\%}&\textbf{11.02\%}&4.54\%&\textbf{3.08\%}&3.38\%&	4.14\%&	\textbf{2.76\%}\\
&Q-Attack(LPA)&	0.62\%&-9.52\%&2.48\%&	\textbf{9.12\%}&2.75\%	&3.27\%	&1.45\%	&-0.08\%\\
&Q-Attack(INF)&	1.69\%&-9.65\%&2.70\%&	3.99\%&\textbf{3.43\%}&	\textbf{4.10\%}&	1.04\%&	0.57\%\\
&Q-Attack(Node2vec+KM)&	-0.69\%&-5.69\%	&2.32\%	&5.25\%&2.35\%	&\textbf{4.59\%}	&1.35\%	&0.71\%\\
\hline
\multirow{6}{*}{Polbooks}&Q-Attack(FN)&\textbf{22.17\%}&2.92\%&2.28\%&4.48\%&2.76\%		&\textbf{3.13\%}	&\textbf{6.29\%}&3.11\%\\
&Q-Attack(SOA)&	4.41\%&\textbf{16.35\%}	&2.46\%&3.74\%	&2.26\%		&2.70\%	&5.32\%	&3.11\% \\
&Q-Attack(LOU)&	3.49\%&4.20\%	&\textbf{17.92\%}&\textbf{5.58\%}&\textbf{3.13\%}	&2.90\%	&\textbf{6.20\%}&\textbf{3.86\%} \\
&Q-Attack(LPA)&	\textbf{8.98\%}	&3.79\%	&1.12\%	&\textbf{5.40\%}&2.38\%	&3.05\%	&4.12\%	&\textbf{3.86\%}\\
&Q-Attack(INF)&	2.45\%&\textbf{6.51\%}&\textbf{5.10\%}&4.67\%	&\textbf{7.48\%}&2.53\%	&4.79\%	&\textbf{4.25\%}\\
&Q-Attack(Node2vec+KM)&1.62\%&2.44\%&2.29\%&	4.38\%&2.42\%&	\textbf{5.82\%}&3.16\%	&2.63\%\\
\hline\hline
\end{tabular}}
\end{table*}

\subsection{Transferability of Q-Attack}
As we can see, Q-Attack is typically based on the modularity $Q$ which depends on certain community detection algorithm. However, there are a large number of community detection algorithms in reality, it's hard for us know which one has been applied. Q-Attack may be of less practicability if the adversarial network obtained according to a given community detection algorithm loses its effectiveness on attacking others. In order to address this concern, we design experiments to testify the transferability of Q-Attack. The relative reductions of modularity Q are presented in Table~\ref{tab:trans}. Again, for each case, we mark the top two best results as bold.

We find that the adversarial networks obtained by Q-Attack on specific community detection algorithms still show considerable attack effects for others in most cases. Another interesting finding is that adversarial networks obtained by taking attacks on Louvain and Infomap show significant transfer attack effects more frequently, while generalized to other algorithms. Meanwhile, Louvain and Infomap seem more robust since reductions of $Q$ on these two algorithms are relatively small in most cases, except taking attacks on these two algorithms directly. The possible explanation for our finding is that adversarial networks obtained by taking attacks on algorithms with better performance will show a stronger transferability. It is consistent with the findings in the researches guided by  Lancichinetti et al.~\cite{lancichinetti2009community} and Yang et al.~\cite{yang2016comparative}, which show Infomap and Louvain have better performance through comparative analysis of different community detection algorithms.

\section{Conclusion}
\label{section5}
In this paper, we propose several strategies, including CDA, DBA and GA based Q-Attack, to attack a number of community detection algorithms for networks. We perform the experiments on four well-known social networks, and find that these attack strategies are effective to make the community detection algorithms fail partly in most cases, in terms of decreasing modularity $Q$ and NMI, by just disturbing a small number of connections. By comparison, Q-Attack behaves much better than CDA and DBA in almost all the considered white-box cases, where the target community detection algorithms are known in advance. Moreover, this attack strategy also presents certain transferability to make effective black-box attacks. That is, the adversarial networks, generated by Q-Attack on particular community detection algorithms, can still be used to effectively attack many others. These results indicate that it is feasible to utilize Q-Attack to protect people's privacy against community detection in reality.

With more and more attentions are focused on the problem of information over-mined in social networks, many works are left to be studied. For instance, here our Q-Attack method only takes modularity $Q$ as the optimization objective due to its simplicity, methods based on multi-objective optimization are also worth studying and may present better attack effects. And for wide applications on lager networks, designing attack strategies of higher efficiency by simplifying the fitness calculation and choosing the optimization algorithms with lower time complexity is one of the research emphases in the future. Besides, the community detection algorithms we have considered in the experiments are mainly used to detect non-overlapping communities, but there are also many overlapping communities in the real world, the attack strategies thus can be extended to such more general situations.

\section*{Acknowledgment}
The authors would like to thank all the members in our IVSN research group in Zhejiang University of Technology for the valuable discussion about the ideas and technical details presented in this paper.

\bibliography{mybibfile}

\begin{thebibliography}{10}
\providecommand{\url}[1]{#1}
\csname url@samestyle\endcsname
\providecommand{\newblock}{\relax}
\providecommand{\bibinfo}[2]{#2}
\providecommand{\BIBentrySTDinterwordspacing}{\spaceskip=0pt\relax}
\providecommand{\BIBentryALTinterwordstretchfactor}{4}
\providecommand{\BIBentryALTinterwordspacing}{\spaceskip=\fontdimen2\font plus
\BIBentryALTinterwordstretchfactor\fontdimen3\font minus
  \fontdimen4\font\relax}
\providecommand{\BIBforeignlanguage}[2]{{%
\expandafter\ifx\csname l@#1\endcsname\relax
\typeout{** WARNING: IEEEtran.bst: No hyphenation pattern has been}%
\typeout{** loaded for the language `#1'. Using the pattern for}%
\typeout{** the default language instead.}%
\else
\language=\csname l@#1\endcsname
\fi
#2}}
\providecommand{\BIBdecl}{\relax}
\BIBdecl

\bibitem{xuan2018social}
Q.~Xuan, Z.-Y. Zhang, C.~Fu, H.-X. Hu, and V.~Filkov, ``Social synchrony on
  complex networks,'' \emph{IEEE transactions on cybernetics}, vol.~48, no.~5,
  pp. 1420--1431, 2018.

\bibitem{xuan2015temporal}
Q.~Xuan, H.~Fang, C.~Fu, and V.~Filkov, ``Temporal motifs reveal collaboration
  patterns in online task-oriented networks,'' \emph{Physical Review E},
  vol.~91, no.~5, p. 052813, 2015.

\bibitem{garcia2018applications}
J.~O. Garcia, A.~Ashourvan, S.~Muldoon, J.~M. Vettel, and D.~S. Bassett,
  ``Applications of community detection techniques to brain graphs: Algorithmic
  considerations and implications for neural function,'' \emph{Proceedings of
  the IEEE}, vol. 106, no.~5, pp. 846--867, 2018.

\bibitem{chen2018robustness}
Z.~Chen, J.~Wu, Y.~Xia, and X.~Zhang, ``Robustness of interdependent power
  grids and communication networks: A complex network perspective,'' \emph{IEEE
  Transactions on Circuits and Systems II: Express Briefs}, vol.~65, no.~1, pp.
  115--119, 2018.

\bibitem{schiavo2010international}
S.~Schiavo, J.~Reyes, and G.~Fagiolo, ``International trade and financial
  integration: a weighted network analysis,'' \emph{Quantitative Finance},
  vol.~10, no.~4, pp. 389--399, 2010.

\bibitem{lancichinetti2009community}
A.~Lancichinetti and S.~Fortunato, ``Community detection algorithms: a
  comparative analysis,'' \emph{Physical review E}, vol.~80, no.~5, p. 056117,
  2009.

\bibitem{fu2018link}
C.~Fu, M.~Zhao, L.~Fan, X.~Chen, J.~Chen, Z.~Wu, Y.~Xia, and Q.~Xuan, ``Link
  weight prediction using supervised learning methods and its application to
  yelp layered network,'' \emph{IEEE Transactions on Knowledge and Data
  Engineering}, 2018.

\bibitem{kipf2016semi}
T.~N. Kipf and M.~Welling, ``Semi-supervised classification with graph
  convolutional networks,'' \emph{arXiv preprint arXiv:1609.02907}, 2016.

\bibitem{zhang2018reconstructing}
H.-F. Zhang, F.~Xu, Z.-K. Bao, and C.~Ma, ``Reconstructing of networks with
  binary-state dynamics via generalized statistical inference,'' \emph{IEEE
  Transactions on Circuits and Systems I: Regular Papers}, 2018.

\bibitem{newman2003structure}
M.~E. Newman, ``The structure and function of complex networks,'' \emph{SIAM
  review}, vol.~45, no.~2, pp. 167--256, 2003.

\bibitem{newman2001structure}
------, ``The structure of scientific collaboration networks,''
  \emph{Proceedings of the national academy of sciences}, vol.~98, no.~2, pp.
  404--409, 2001.

\bibitem{gleiser2003community}
P.~M. Gleiser and L.~Danon, ``Community structure in jazz,'' \emph{Advances in
  complex systems}, vol.~6, no.~04, pp. 565--573, 2003.

\bibitem{abbasi2007survey}
A.~A. Abbasi and M.~Younis, ``A survey on clustering algorithms for wireless
  sensor networks,'' \emph{Computer communications}, vol.~30, no. 14-15, pp.
  2826--2841, 2007.

\bibitem{fortunato2016community}
S.~Fortunato and D.~Hric, ``Community detection in networks: A user guide,''
  \emph{Physics Reports}, vol. 659, pp. 1--44, 2016.

\bibitem{xuan2018modern}
Q.~Xuan, M.~Zhou, Z.-Y. Zhang, C.~Fu, Y.~Xiang, Z.~Wu, and V.~Filkov, ``Modern
  food foraging patterns: Geography and cuisine choices of restaurant patrons
  on yelp,'' \emph{IEEE Transactions on Computational Social Systems}, vol.~5,
  no.~2, pp. 508--517, 2018.

\bibitem{nagaraja2010impact}
S.~Nagaraja, ``The impact of unlinkability on adversarial community detection:
  effects and countermeasures,'' in \emph{International Symposium on Privacy
  Enhancing Technologies Symposium}.\hskip 1em plus 0.5em minus 0.4em\relax
  Springer, 2010, pp. 253--272.

\bibitem{waniek2018hiding}
M.~Waniek, T.~P. Michalak, M.~J. Wooldridge, and T.~Rahwan, ``Hiding
  individuals and communities in a social network,'' \emph{Nature Human
  Behaviour}, vol.~2, no.~2, p. 139, 2018.

\bibitem{fionda2018community}
V.~Fionda and G.~Pirro, ``Community deception or: How to stop fearing community
  detection algorithms,'' \emph{IEEE Transactions on Knowledge and Data
  Engineering}, vol.~30, no.~4, pp. 660--673, 2018.

\bibitem{holme2002attack}
P.~Holme, B.~J. Kim, C.~N. Yoon, and S.~K. Han, ``Attack vulnerability of
  complex networks,'' \emph{Physical review E}, vol.~65, no.~5, p. 056109,
  2002.

\bibitem{tasgin2007community}
M.~Tasgin, A.~Herdagdelen, and H.~Bingol, ``Community detection in complex
  networks using genetic algorithms,'' \emph{arXiv preprint arXiv:0711.0491},
  2007.

\bibitem{liu2011application}
H.~Liu, X.-B. Hu, S.~Yang, K.~Zhang, and E.~Di~Paolo, ``Application of complex
  network theory and genetic algorithm in airline route networks,''
  \emph{Transportation Research Record}, vol. 2214, no.~1, pp. 50--58, 2011.

\bibitem{wang2012community}
S.~Wang, H.~Zou, Q.~Sun, X.~Zhu, and F.~Yang, ``Community detection via
  improved genetic algorithm in complex network,'' \emph{Information Technology
  Journal}, vol.~11, no.~3, pp. 384--387, 2012.

\bibitem{girvan2002community}
M.~Girvan and M.~E. Newman, ``Community structure in social and biological
  networks,'' \emph{Proceedings of the national academy of sciences}, vol.~99,
  no.~12, pp. 7821--7826, 2002.

\bibitem{tyler2005mail}
J.~R. Tyler, D.~M. Wilkinson, and B.~A. Huberman, ``E-mail as spectroscopy:
  Automated discovery of community structure within organizations,'' \emph{The
  Information Society}, vol.~21, no.~2, pp. 143--153, 2005.

\bibitem{radicchi2004defining}
F.~Radicchi, C.~Castellano, F.~Cecconi, V.~Loreto, and D.~Parisi, ``Defining
  and identifying communities in networks,'' \emph{Proceedings of the National
  Academy of Sciences}, vol. 101, no.~9, pp. 2658--2663, 2004.

\bibitem{newman2004fast}
M.~E. Newman, ``Fast algorithm for detecting community structure in networks,''
  \emph{Physical review E}, vol.~69, no.~6, p. 066133, 2004.

\bibitem{clauset2004finding}
A.~Clauset, M.~E. Newman, and C.~Moore, ``Finding community structure in very
  large networks,'' \emph{Physical review E}, vol.~70, no.~6, p. 066111, 2004.

\bibitem{blondel2008fast}
V.~D. Blondel, J.-L. Guillaume, R.~Lambiotte, and E.~Lefebvre, ``Fast unfolding
  of communities in large networks,'' \emph{Journal of statistical mechanics:
  theory and experiment}, vol. 2008, no.~10, p. P10008, 2008.

\bibitem{newman2006modularity}
M.~E. Newman, ``Modularity and community structure in networks,''
  \emph{Proceedings of the national academy of sciences}, vol. 103, no.~23, pp.
  8577--8582, 2006.

\bibitem{newman2006finding}
------, ``Finding community structure in networks using the eigenvectors of
  matrices,'' \emph{Physical review E}, vol.~74, no.~3, p. 036104, 2006.

\bibitem{raghavan2007near}
U.~N. Raghavan, R.~Albert, and S.~Kumara, ``Near linear time algorithm to
  detect community structures in large-scale networks,'' \emph{Physical review
  E}, vol.~76, no.~3, p. 036106, 2007.

\bibitem{rosvall2008maps}
M.~Rosvall and C.~T. Bergstrom, ``Maps of random walks on complex networks
  reveal community structure,'' \emph{Proceedings of the National Academy of
  Sciences}, vol. 105, no.~4, pp. 1118--1123, 2008.

\bibitem{ronhovde2009multiresolution}
P.~Ronhovde and Z.~Nussinov, ``Multiresolution community detection for
  megascale networks by information-based replica correlations,''
  \emph{Physical Review E}, vol.~80, no.~1, p. 016109, 2009.

\bibitem{bellingeri2014efficiency}
M.~Bellingeri, D.~Cassi, and S.~Vincenzi, ``Efficiency of attack strategies on
  complex model and real-world networks,'' \emph{Physica A: Statistical
  Mechanics and its Applications}, vol. 414, pp. 174--180, 2014.

\bibitem{karrer2008robustness}
B.~Karrer, E.~Levina, and M.~E. Newman, ``Robustness of community structure in
  networks,'' \emph{Physical Review E}, vol.~77, no.~4, p. 046119, 2008.

\bibitem{yu2018target}
S.~Yu, M.~Zhao, C.~Fu, H.~Huang, X.~Shu, Q.~Xuan, and G.~Chen, ``Target defense
  against link-prediction-based attacks via evolutionary perturbations,''
  \emph{arXiv preprint arXiv:1809.05912}, 2018.

\bibitem{dai2018adversarial}
H.~Dai, H.~Li, T.~Tian, X.~Huang, L.~Wang, J.~Zhu, and L.~Song, ``Adversarial
  attack on graph structured data,'' \emph{arXiv preprint arXiv:1806.02371},
  2018.

\bibitem{zugner2018adversarial}
D.~Z{\"u}gner, A.~Akbarnejad, and S.~G{\"u}nnemann, ``Adversarial attacks on
  neural networks for graph data,'' in \emph{Proceedings of the 24th ACM SIGKDD
  International Conference on Knowledge Discovery \& Data Mining}.\hskip 1em
  plus 0.5em minus 0.4em\relax ACM, 2018, pp. 2847--2856.

\bibitem{bojchevski2018adversarial}
A.~Bojchevski and S.~G\"{u}nnemann, ``Adversarial attacks on node embeddings,''
  \emph{arXiv preprint arXiv:1809.01093}, 2018.

\bibitem{chen2018fast}
J.~Chen, Y.~Wu, X.~Xu, Y.~Chen, H.~Zheng, and Q.~Xuan, ``Fast gradient attack
  on network embedding,'' \emph{arXiv preprint arXiv:1809.02797}, 2018.

\bibitem{barabasi1999emergence}
A.-L. Barab{\'a}si and R.~Albert, ``Emergence of scaling in random networks,''
  \emph{science}, vol. 286, no. 5439, pp. 509--512, 1999.

\bibitem{holland1975adaptation}
J.~Holland, ``Adaptation in natural and artificial systems: an introductory
  analysis with application to biology,'' \emph{Control and artificial
  intelligence}, 1975.

\bibitem{kirkpatrick1983optimization}
S.~Kirkpatrick, C.~D. Gelatt, and M.~P. Vecchi, ``Optimization by simulated
  annealing,'' \emph{science}, vol. 220, no. 4598, pp. 671--680, 1983.

\bibitem{eberhart1995new}
R.~Eberhart and J.~Kennedy, ``A new optimizer using particle swarm theory,'' in
  \emph{Micro Machine and Human Science, 1995. MHS'95., Proceedings of the
  Sixth International Symposium on}.\hskip 1em plus 0.5em minus 0.4em\relax
  IEEE, 1995, pp. 39--43.

\bibitem{zachary1977information}
W.~W. Zachary, ``An information flow model for conflict and fission in small
  groups,'' \emph{Journal of anthropological research}, vol.~33, no.~4, pp.
  452--473, 1977.

\bibitem{lusseau2003bottlenose}
D.~Lusseau, K.~Schneider, O.~J. Boisseau, P.~Haase, E.~Slooten, and S.~M.
  Dawson, ``The bottlenose dolphin community of doubtful sound features a large
  proportion of long-lasting associations,'' \emph{Behavioral Ecology and
  Sociobiology}, vol.~54, no.~4, pp. 396--405, 2003.

\bibitem{grover2016node2vec}
A.~Grover and J.~Leskovec, ``node2vec: Scalable feature learning for
  networks,'' in \emph{Proceedings of the 22nd ACM SIGKDD international
  conference on Knowledge discovery and data mining}.\hskip 1em plus 0.5em
  minus 0.4em\relax ACM, 2016, pp. 855--864.

\bibitem{newman2004finding}
M.~E. Newman and M.~Girvan, ``Finding and evaluating community structure in
  networks,'' \emph{Physical review E}, vol.~69, no.~2, p. 026113, 2004.

\bibitem{danon2005comparing}
L.~Danon, A.~Diaz-Guilera, J.~Duch, and A.~Arenas, ``Comparing community
  structure identification,'' \emph{Journal of Statistical Mechanics: Theory
  and Experiment}, vol. 2005, no.~09, p. P09008, 2005.

\bibitem{yang2016comparative}
Z.~Yang, R.~Algesheimer, and C.~J. Tessone, ``A comparative analysis of
  community detection algorithms on artificial networks,'' \emph{Scientific
  reports}, vol.~6, p. 30750, 2016.

\end{thebibliography}

\end{document}